\documentclass[12pt,preprint]{aastex}
\usepackage{emulateapj5}

\slugcomment{ApJ Letters, in press}

\shorttitle{Clustering Properties of Galaxies at $z\sim 4$}
\shortauthors{Ouchi et al.}

\begin{document}


\title{Clustering Properties of Galaxies at $z\sim4$\\
in the Subaru/XMM Deep Survey Field \altaffilmark{1} }


\author{Masami Ouchi        \altaffilmark{2},
	Kazuhiro Shimasaku  \altaffilmark{2,3},
	Sadanori Okamura    \altaffilmark{2,3},\\
	Mamoru Doi          \altaffilmark{4},
	Hisanori Furusawa   \altaffilmark{2},
	Masaru Hamabe       \altaffilmark{5},
	Masahiko Kimura     \altaffilmark{6},\\
	Yutaka Komiyama     \altaffilmark{7},
	Masayuki Miyazaki   \altaffilmark{2},
	Satoshi Miyazaki    \altaffilmark{8},
	Fumiaki Nakata      \altaffilmark{2},\\
	Maki Sekiguchi      \altaffilmark{6},
	Masafumi Yagi       \altaffilmark{8}, and
	Naoki Yasuda        \altaffilmark{8}
	}

\email{ouchi@astron.s.u-tokyo.ac.jp}


\altaffiltext{1}{Based on data collected at 
	Subaru Telescope, which is operated by 
	the National Astronomical Observatory of Japan.}
\altaffiltext{2}{Department of Astronomy, School of Science,
        University of Tokyo, Tokyo 113-0033, Japan}
\altaffiltext{3}{Research center for the Early Universe, School of Science,
        University of Tokyo, Tokyo 113-0033, Japan}
\altaffiltext{4}{Institute of Astronomy, School of Science, 
	University of Tokyo, Mitaka, Tokyo 181-0015, Japan}
\altaffiltext{5}{Department of Mathematical and Physical Sciences,
	Faculty of Science, Japan Women's University, Tokyo 112-8681, Japan}
\altaffiltext{6}{Institute for Cosmic Ray Research, 
	University of Tokyo, Kashiwa, Chiba 277-8582}
\altaffiltext{7}{Subaru Telescope, National Astronomical Observatory, 
	650 N.A'ohoku Place, Hilo, HI 96720, USA}
\altaffiltext{8}{National Astronomical Observatory, 
	Mitaka, Tokyo 181-8588, Japan}


\begin{abstract}
We study the clustering properties of about 1200 $z\sim 4$
Lyman Break Galaxy (LBG) candidates with $i'<26$ 
which are selected by color from deep $BRi'$ imaging 
data of a 618 arcmin$^2$ area in the Subaru/XMM-Newton 
Deep Field taken with Subaru Prime Focus Camera.
The contamination and completeness of our LBG sample 
are evaluated, on the basis of 
the Hubble Deep Field North (HDFN) objects, 
to be 17\% and 45\%, respectively.
We derive the angular correlation function 
over $\theta = 2'' - 1000''$, and find that it is fitted fairly well 
by a power law, $\omega (\theta)=A_\omega \theta^{-0.8}$, 
with $A_\omega = 0.71 \pm 0.26$.
We then calculate the correlation length $r_0$ 
(in comoving units) of the two-point spatial correlation 
function $\xi(r) = (r/r_0)^{-1.8}$ from $A_\omega$ 
using the redshift distribution of LBGs derived from the HDFN, 
and find $r_0=2.7^{+0.5}_{-0.6}$ $h^{-1}$ Mpc 
in a $\Lambda$-dominated universe 
($\Omega_m=0.3$ and $\Omega_\Lambda=0.7$).
This is twice larger than the correlation length 
of the dark matter at $z\simeq 4$ 
predicted from 
an analytic model by Peacock \& Dodds
but about twice smaller than that of bright galaxies
predicted by 
a semi-analytic model of Baugh et al.
We find an excess of $\omega(\theta)$ on small scales 
($\theta \lesssim 5''$) departing from the power law fit 
over 3 $\sigma$ significance levels.
Interpreting this as due to galaxy mergers,
we estimate the fraction of galaxies undergoing mergers
in our LBG sample to be $3.0 \pm 0.9\%$, which 
is significantly smaller than those of galaxies at intermediate 
redshifts.

\end{abstract}





\section{INTRODUCTION}
\label{sec:introduction}

Studies of clustering properties of galaxies 
at high redshifts 
are essential for understanding galaxy formation and its 
relationship to structure formation.
The most efficient way to select high redshift ($z > 2.5$) 
galaxies is the Lyman break technique.
High redshift galaxies are isolated in a two-color 
plane using their UV continuum properties, 
and galaxies selected in this way are called 
Lyman break galaxies (LBGs).
\citet{giavalisco1998} studied clustering properties 
of LBGs 
on the basis of a large sample for bright ($R<25.5$) LBGs 
lying around $z=3$ selected by $U_nGR$ colors.
They found that the spatial distribution of $z=3$ LBGs
is strongly biased relative to the expectations for the
dark matter in Cold Dark Matter models
with a linear bias of 4.5 predicted for
an Einstein-de Sitter cosmology 
(see also \citealt{adelberger1998}).
In a subsequent paper, \citeauthor{giavalisco2001}
(\citeyear{giavalisco2001}; GD01) 
found the clustering amplitude 
of $z\sim 3$ LBGs to depend on their rest-frame 
ultraviolet luminosity,
with fainter galaxies being less strongly clustered.
\citet{arnouts1999} measured the correlation length of 
faint galaxies in the Hubble Deep Field North (HDFN) 
over the redshift range $0<z<4$,
though errors in their measurements are large 
due to small statistics.

The aim of this paper is to extend the studies
on galaxy clustering to $z\sim 4$ LBGs 
using a large sample of about 1200 LBG candidates 
identified by the two color ($B-R$ vs $R-i'$) 
selection technique. 
This is the largest sample of $z\sim 4$ LBGs 
in a contiguous area obtained to date, 
and thus enables detailed studies of clustering properties 
of galaxies at the highest redshift.
Throughout this paper, magnitudes are in the AB system.

\section{OBSERVATIONS AND DATA REDUCTION}
\label{sec:observations_and_data_reduction}
Deep and wide-field $B$-,$V$-,$R$- and $i'$-band imaging data 
of a central $30'\times 24'$ area in the 
Subaru/XMM-Newton Deep Survey Field 
($2^h 18^m 00^s$,$-5^\circ 12 ' 00''$[J2000]) were taken with 
Subaru Prime Focus Camera (Suprime-Cam; \citealt{miyazaki1998})
during the commissioning observing runs on November 24-27, 2000. 
The present work is based on the $B$, $R$, and $i'$ data.
The individual CCD data were reduced and
combined using IRAF and the mosaic-CCD data reduction software 
developed by us (Yagi 1998).
The combined images for individual bands were aligned and 
smoothed with Gaussian kernels to match their seeing sizes. 
The final images cover a contiguous
618 arcmin$^2$ area with a PSF FWHM of $0.''98$.
The net exposure times of the final images are 
177, 58, and 45 minutes for $B$, $R$, and $i'$, 
respectively. 
The limiting magnitudes are $B=27.6$, $R=26.5$, and 
$i'=26.2$ for a 3$\sigma$ detection on a
$2''$ diameter aperture.
Source detection and photometry are performed using
SExtractor version 2.1.6 (\citealt{bertin1996}).
The $i'$-band frame is chosen to detect objects, 
and we limit the object catalog to 
$i' \leq 26$, in order to provide 
a reasonable level of photometric completeness.

\section{SELECTION OF $z \sim 4$ LYMAN BREAK GALAXIES}
\label{sec:selection_sample}
Our catalog contains 42,557 objects 
with $i'\leq 26.0$ in total.
On the basis of expectations
from GISSEL96 (\citealt{bruzual1993}) population synthesis models
\footnote{The model parameters are chosen to match the observed
colors of $z=3$ galaxies: Salpeter IMF, $Z_{metal}=0.2Z_\odot$,
and an age of 10 Myr for an instantanious burst 
(\citealt{sawicki1998}).}
with dust attenuation of $E(B-V) = 0-0.5$ 
assuming \citeauthor{calzetti1997}'s (\citeyear{calzetti1997}) 
extinction law, we define the photometric selection criteria
for galaxies at $z \sim 4$ as
\begin{equation}
B-R > 2.1, 
R-i' < 0.5, 
B-R > 5.4(R-i')+0.9 \\
\label{eq:lbgselection}
\end{equation}
\noindent
We estimate from the models that the expected redshift distribution of 
galaxies satisfying eq. (\ref{eq:lbgselection}) 
is $z=3.8\pm0.5$ (\citealt{ouchi2001}).

\begin{figure*}
\epsscale{0.8}
\plotone{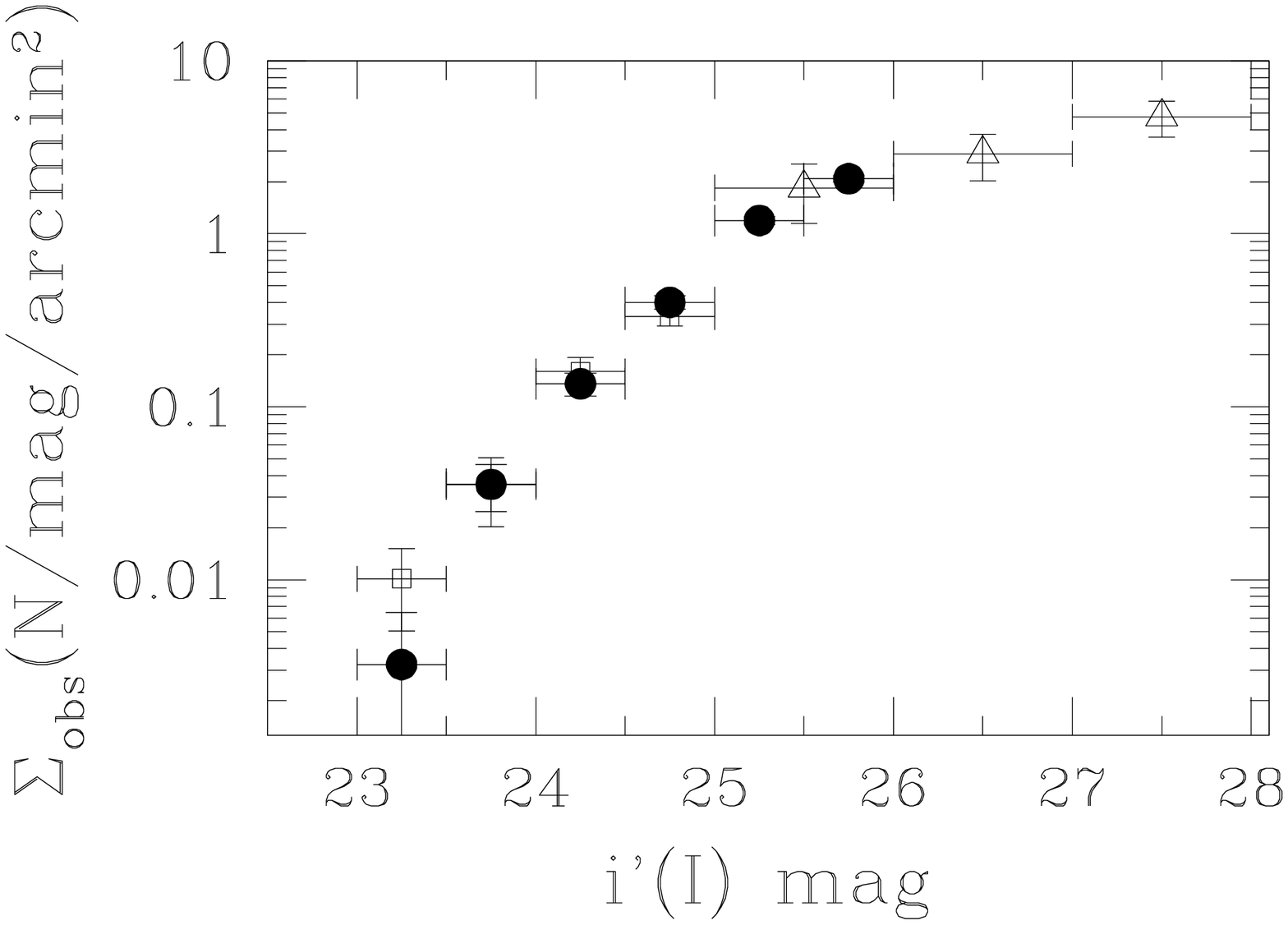}
\caption{Observed surface densities of  
    $z \sim 4$ LBGs in $i'$ (or $I$) magnitude 
    without correction for incompleteness of detections.
    Filled circles show our 1192 objects. Open squares are
    similarly color-selected objects using $GRI$ 
    by \citet{steidel1999}.
    Triangles indicate objects in the HDFN selected by us 
    using eq.(\ref{eq:lbgselection}), whose magnitudes 
    have been transformed to 
    our $BR{\it i'}$ magnitudes following \citet{ouchi2001}. 
    \label{fig:m_sigma}}
\end{figure*}

There are a total of 1,192 objects which meet the criteria. 
Figure \ref{fig:m_sigma} shows the number counts of the objects, 
without correction for incompleteness of detections, 
against ${\it i'}$ (filled circles).
It is found that our counts are consistent with those 
given in \citet{steidel1999}, who selected $z\sim 4$ LBGs 
using $GRI$ color, 
and that the faint end of our counts matches the counts derived 
from the HDFN.
For the following analyses, we define three LBG samples; 
(1) the whole sample containing 1,192 objects,
(2) the sample for bright ($i'<25.5$) objects,
and (3) the sample for faint ($25.5<i'<26$) objects.

We have estimated the contamination and 
completeness of our LBG sample by Monte Carlo 
simulations, generating artificial galaxies
using the HDFN objects for which 
magnitudes, colors, and photometric redshifts are 
given in \citet{fernandez-soto1999},
and distributing them on our original images (\citealt{ouchi2001}).
The contamination is defined, 
for the detected simulated objects, as the ratio of 
low redshift ($z<3.3$) objects meeting eq.(\ref{eq:lbgselection}), 
to all objects satisfying eq.(\ref{eq:lbgselection}).
The completeness is defined as the ratio of $i'<26$ simulated objects 
passing our detection threshold 
{\it and} satisfying eq.(\ref{eq:lbgselection}), 
to all (detected $+$ undetected) $i'<26$ simulated objects.
We find from the simulations that the completeness 
is $45\%$, $51\%$, and $40\%$ for the whole, bright, and faint 
samples, respectively. 
Similarly, the contamination is
$17\%$, $19\%$, and $14\%$, respectively.
Note that the completeness defined here is a combination 
of the detection completeness and the completeness of color 
selection.
Our color criteria effectively sample HDFN galaxies at $3.3<z<4.3$ 
with a selection completeness of 89\%. 
Namely, 89\% of $z\simeq 4$ galaxies pass eq.(\ref{eq:lbgselection}) 
once they are detected.
We adopt for the redshift distribution $N(z)$
of LBGs in our sample the one derived from
the simulations taking account of the
contamination and completeness.


\section{ANGULAR CORRELATION FUNCTION}
\label{sec:angular_correlation_function}

\begin{figure}
\plotone{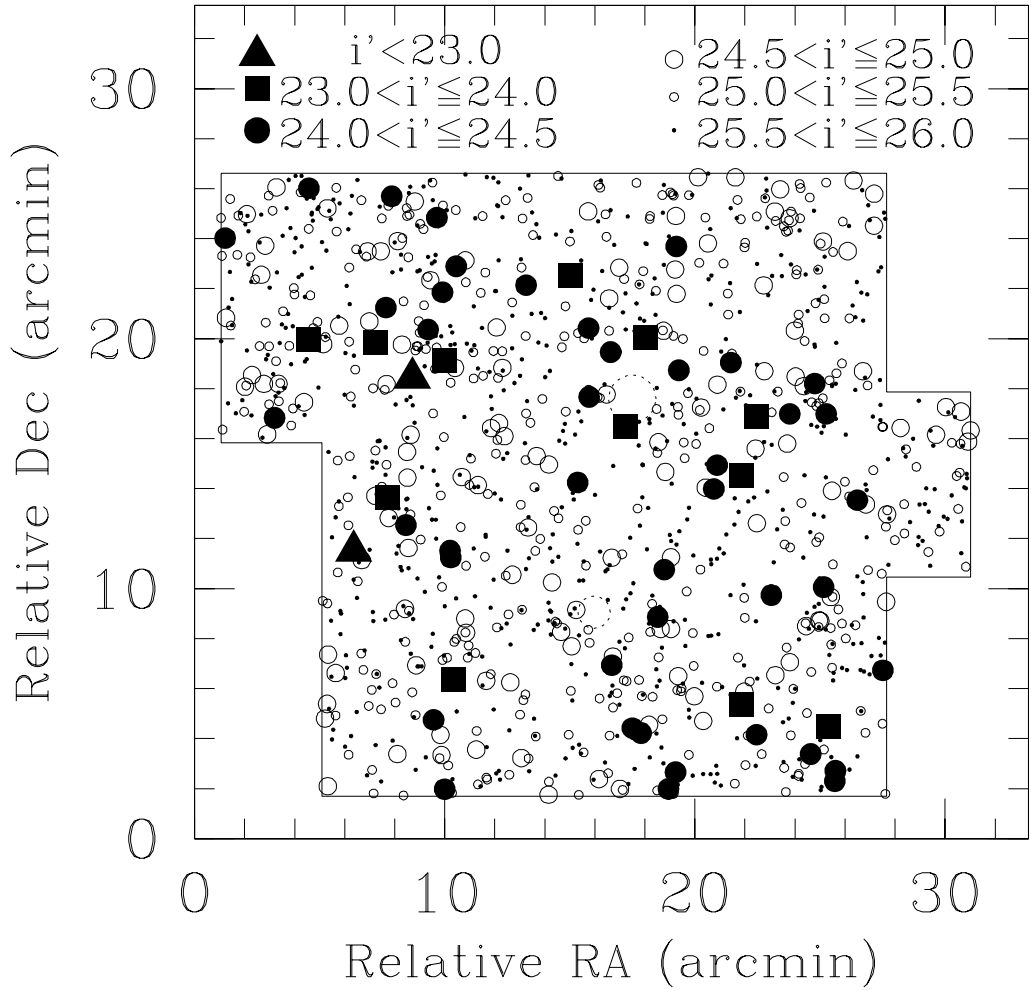}
\caption{Sky distribution of our $z \sim 4$ LBG candidates.
    Different symbols correspond to different magnitude bins 
    defined in the panel. 
    Masked regions to avoid effects 
    of bright stars are shown by dashed circles.
    \label{fig:lbgdistribution}}
\end{figure}

Figure \ref{fig:lbgdistribution} shows the sky distribution 
of $z\sim 4$ LBGs in our sample.
We find in this figure somewhat an inhomogeneous distribution 
of LBGs, especially for bright ($i'<24.5$) ones.
This is not due to the inhomogeneity of 
detection completeness over the image,
because the Monte Carlo simulations show that differences in 
detection completeness among small ($1.'7 \times 1.'7$) areas 
in the image are less than $10\%$, and that 
the detection completeness does not correlate with 
the distribution of LBGs.

We derive the angular two-point correlation function,
$\omega$($\theta$), using the estimator defined by \citet{landy1993}, 
$
\omega_{obs}(\theta)
  = [DD(\theta)-2DR(\theta)+RR(\theta)]/RR(\theta),
$
where $DD(\theta)$, $DR(\theta)$, and $RR(\theta)$ are numbers of
galaxy-galaxy, galaxy-random, and random-random pairs normalized by
the total number of pairs in each of the three samples.
The random sample is composed of 200,000 sources with the same
geometrical constraints of the data sample. 
The formal error \footnote{The formal error does not 
include sample variance, which is
caused from field-to-field variations.}
 in $\omega$($\theta$) is assigned by
$\sigma_\omega = \sqrt{[1+\omega_{obs}(\theta)]/RR}$.
The real correlation function $\omega(\theta)$ is
offset by the integral constant ($IC$: Groth \& Peebles 1977);
$\omega(\theta)= \omega_{obs}(\theta)+IC$
,where $IC$ of the whole sample is calculated to be 0.0045.

The resulting angular correlation function for the
whole sample is shown in 
Figure \ref{fig:resultacorr_all2} after the 
application of $IC$. 
We fit a power law, 
$\omega(\theta)=A_\omega \theta^{-\beta}$, 
to the data points, and find the slope 
$\beta = 0.6^{+0.6}_{-0.4}$, which is 
consistent with those for nearby galaxies, $\beta = 0.7-0.8$, 
though the errors in our estimate are large.
The best fit values of $A_\omega$ for $\beta \equiv 0.8$
are summarized in Table \ref{tab:sample}.

The two-point angular correlation function is related
to the spatial correlation function 
$\xi(r) = (r/r_0)^{-\gamma}$ by an integral equation,
the Limber transformation
(\citealt{peebles1980};\citealt{efstathiou1991}). 
\begin{equation}
      A_{\omega}= C r_0^\gamma
      \int_{0}^{\infty} F(z) D_\theta^{1-\gamma}(z)N(z)^2g(z) dz 
      \left[ \int_{0}^{\infty} N(z) dz \right]^{-2}
 \label{eq:limber}
\end{equation}

 \noindent
 where 
$F(z)$ \footnote{
Assuming that the clustering pattern 
is fixed in comoving coordinates in the
redshift range of our sample, 
we take the functional form,
$F(z)={(1+z)/(1+3.8)}^{-(3+\epsilon)}$,
where $\epsilon=-1.2$. 
The effect of the change in $\epsilon$ over
$0<\epsilon<-3$ on $r_0$ is, however, very small.}
describes the redshift dependence of $\xi(r)$, 
 $D_\theta(z)$is the angular diameter distance, 
 $
 g(z)=\frac{H_0}{c}[(1+z)^2(1+\Omega_m z 
     + \Omega_\Lambda((1+z)^{-2}-1))^{1/2}],
 $
 and $C$ is a numerical constant 
 $C=\sqrt{\pi} \frac{\Gamma[(\gamma -1)/2]}{\Gamma(\gamma/2)}$.
The slope $\beta$ is related by $\beta=\gamma-1$. 
The Limber transformation requires the redshift distribution 
$N(z)$ of LBGs (see \S \ref{sec:selection_sample}).
We adopt the one derived from the 
simulations based on the HDFN galaxies 
(see \S \ref{sec:selection_sample}).
The obtained values of $r_0$ are presented in Table \ref{tab:sample}
for three cosmologies, 
$(\Omega_m,\Omega_\Lambda)=(0.3, 0.7), (0.3, 0)$,and $(1, 0)$.
Those correlation lengths correspond to projected
angular scales of $100''-110''$ at $z=3.8$. 
These scales fall just outside of 
the range where the power-law fitting is excellent 
($10''<\theta<100''$), implying that $r_0$ is reliably 
measured from the fit.
Varying $N(z)$ in a reasonable range changes the results 
only slightly; 
for example, a tophat distribution over $3.3<z<4.3$ returns $8\%$
larger values, and adopting a narrower distribution, $3.4<z<4.2$, 
decreases $r_0$ by $11\%$.
The random contaminant objects dilute
the amplitudes of angular correlation function,
$A_\omega$, by a factor up to $(1-f)^2$, 
where $f$ is the contamination (see \S \ref{sec:selection_sample}).
The contamination correction increases $r_0$ by 
about 20\% at most.
The effect of field-to-field variations in our sample is 
probably modest, since 
our sample probes a large comoving volume, 
$36 \times 36 \times 520 = 6.7\times 10^5 h^{-3}$ Mpc$^3$ 
for $\Omega_m=0.3$ and $\Omega_\Lambda=0.7$ cosmology.
Observations of other fields and analyses of N-body simulations 
for structure formation will be useful for evaluation of 
the effect of field variation.

\section{DISCUSSION \& CONCLUSIONS}
\label{sec:discussion}

\begin{figure}
\plotone{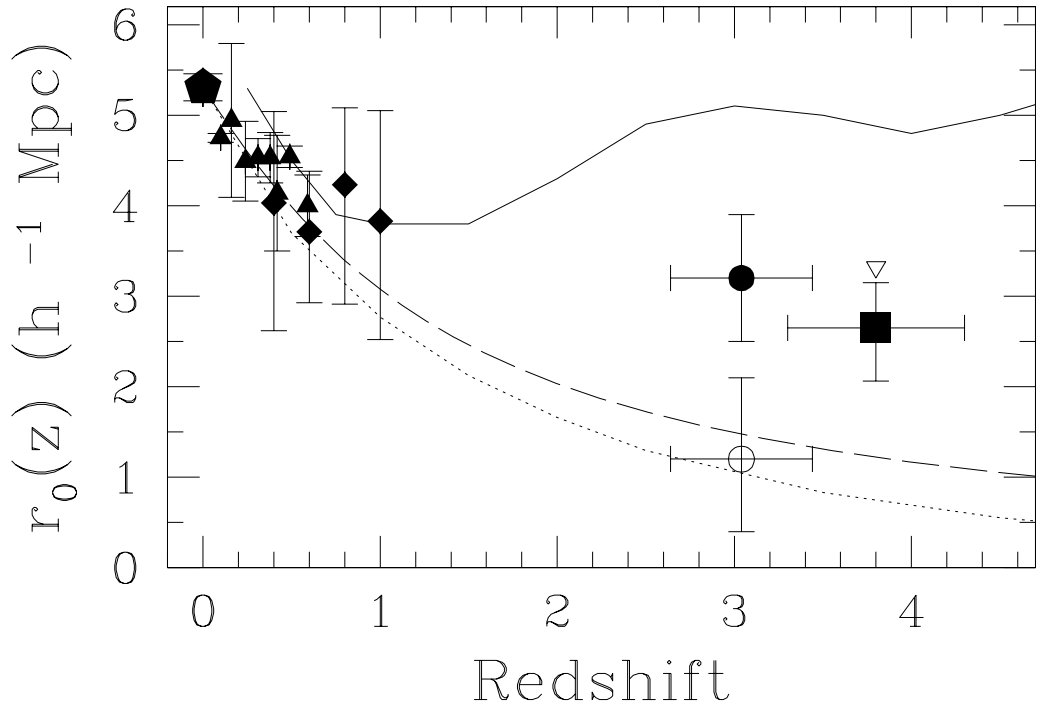}
\caption{Correlation lengths at various redshifts 
  in comoving units. 
  The filled square indicates the value for our whole sample, 
  while filled and open circles are those of
  $z\sim3$ LBGs with $R<25.5$ and $R<27$ (GD01). 
  The open inverse triangle shows the upper limit value for
  our whole sample, when the sample contamination maxmumly 
  dilutes the angular correlation function.
  Filled pentagon, triangles, and diamonds represent 
  measurements for 
  nearby and intermediate-redshift bright galaxies 
  (\citealt{loveday1995}, \citealt{carlberg2000}, \citealt{brunner2000}). 
  The dashed line shows the prediction of linear theory 
  normalized to $r_0(0)=5.3 h^{-1}$ Mpc. 
  Dotted and solid
  lines are $r_0$ of dark matter and galaxies with
  $R <$25.5 (including dust extinction) predicted by
  a semi-analytic model (\citealt{baugh1999}).
  \label{fig:zr0_lambda}}
\end{figure}

%
%
Figure \ref{fig:zr0_lambda} shows
the observed correlation length $r_0$ (in comoving units)
of galaxies at $\left< z \right> = 3.8$ in our sample,
together with those at various redshifts between $z=0$ and $3$.
Two data points at $z=3$ are the measurements by GD01
for bright ($R < 25.5$) LBGs and
faint ($R < 27$) LBGs.
It is found that the $r_0$ for our whole sample
is only slightly smaller than that for GD01's bright LBGs at $z=3$
(
The difference in the mean
brightness in our three samples is less than 1 mag,
and the absolute magnitudes probed by our samples
correspond roughly to those probed by GD01's bright sample).
This
suggests that the clustering amplitude does not
change significantly from $z=3$ to 3.8, with a possible
decrease within $\sim 50\%$.
GD01 found strong luminosity dependence
of $r_0$ for $z=3$ LBGs as shown in Figure \ref{fig:zr0_lambda}.
Our sample does not contain as faint LBGs as they studied,
but we also find possible dependence on luminosity
in our sample, with fainter LBGs having a smaller $r_0$.
\citet{arnouts1999} obtained a large value 
$r_0=(4.28 \pm 1.69) h^{-1}$ Mpc for $z\sim 4$ galaxies in the HDFN.
This is the opposite sense since the galaxies in the HDFN are 
fainter than those in our sample.
However, the number of galaxies they used is only 35. 

The $r_0$ values of LBGs at $z=3-4$ are about half that of 
the nearby galaxies. 
The dotted and solid lines in Figure \ref{fig:zr0_lambda} 
present, respectively, the evolution of $r_0$ 
for the dark matter predicted by an analytic model 
and {\lq}galaxies{\rq} of $R<25.5$ predicted by
a semi-analytic model for galaxy formation 
in a cold dark matter universe with $\Omega_m=0.3$,
$\Omega_\Lambda=0.7$, and $\sigma_8=0.94$ (\citealt{baugh1999}). 
For the reader's guide, we also show by the dashed line 
the $r_0$ of the dark matter 
predicted by linear theory normalized to $r_0 (0)=5.3 h^{-1}$ Mpc. 
The $r_0$ of the observed LBGs at $z=3.8$ is found to be 
$2.9$ times larger than that for the dark matter 
given by Baugh et al. (1999) 
(In practice, Baugh et al. adopted for the
dark matter correlation function the analytic model of 
\citealt{peacock1996}.)
Note that if one compares predicted correlation lengths 
for galaxies with observations, predictions should be made 
based on selection criteria of galaxies which are similar to 
those adopted in the observations.
If we interpret this as biasing of galaxy distribution, 
which is expressed as $\xi_{gal}(r)=b^2 \xi_{matter}(r)$,
we obtain a linear bias $b \sim 2.6$, 
which is similar to that found by GD01 for bright $z=3$ LBGs.
On the other hand, the semi-analytic model 
predicts much stronger clustering for galaxies  
(Similar predictions are made by, eg, \citealt{kauffmann1999}
and \citealt{yoshikawa2001}).
The predicted value at $z=3.8$ is 
twice as large as observed in this study.
Further discussion requires a detailed comparison of galaxy 
properties, including selection criteria, 
between their simulated galaxies and the LBGs in our sample.

\begin{figure}
\plotone{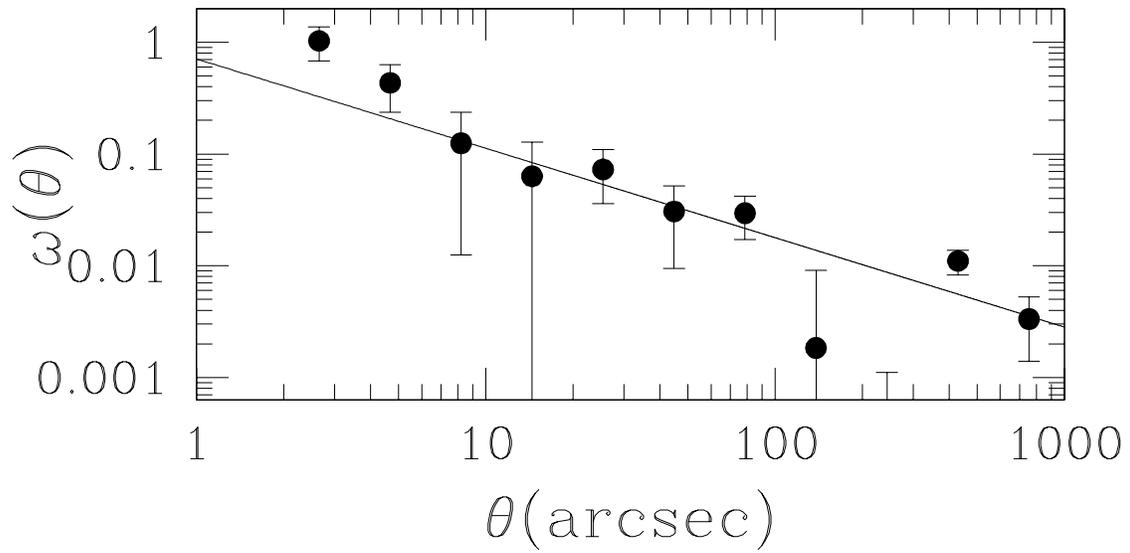}
\caption{Angular correlation function for the whole sample.
    The solid line shows the best fit power law with
    $\omega(\theta)=A_\omega \theta^{-0.8}$.
    \label{fig:resultacorr_all2} }
\end{figure}

In Figure \ref{fig:resultacorr_all2},
we find an excess 
\footnote{
A visual inspection find that all the galaxies contributing to 
the excess are real; no artificial object such as haloes of 
bright stars is found to be included in them.
}
of $\omega(\theta)$ at small scales
($\theta \lesssim 5''$),
over 3 $\sigma$ significance levels, relative to
the best fit power law. 
This may be caused from various
effects, for example, galaxy-galaxy mergers,
two galaxies within
a common dark matter halo, and/or field variance.
If we interpret this as galaxy-galaxy mergers of those
$i'<26$ LBGs, which are classified as bright LBGs
with $L_{1700\AA} \gtrsim L_{1700\AA}^*$ (\citealt{ouchi2001}),
we calculate, following \citet{roche1999}, 
the fraction of galaxies undergoing mergers as
$f_{pair} = 3.0 \pm 0.9\%$ 
from the observed and expected numbers of galaxy pairs 
with a separation of $1.''5 < \theta < 4''$ 
(projected physical separation corresponds to
7.5 to 20 $h^{-1}$kpc at $z=3.8$), 
where the expected number is calculated from 
an integration over $1.''5 < \theta < 4''$ 
of the best fit power law derived for 
the data at $ \theta > 4''$.
This $f_{pair}$ value is significantly smaller than 
those for 
intermediate-redshift ($z \sim 0.3$) galaxies whose 
$f_{pair}$ at $\lesssim$ 20 $h^{-1}$kpc is $5-15\%$  
(\citealt{carlberg1994}; \citealt{infante1996}; \citealt{roche1999}).
This may imply that merger rates for massive
galaxies are lower at $z\sim 4$ than at present,
if we assume that bright LBGs are massive.
However, $i'$-band magnitudes (rest-frame UV
for $z\sim 4$LBGs) are sensitive 
to on-going star-formation activities, and
those luminosities reflect intensity of
current star-formation. Observing those
LBGs in near-IR bands and estimating
their mass are needed to address this issue.


\acknowledgments
We thank the anonymous referee for detailed comments
which improved this manuscript.
We would like to thank the Subaru Telescope staff
for their invaluable help in commisioning the Suprime-Cam
that makes these difficult observations possible.
M. Ouchi, H. Furusawa, and F. Nakata acknowledge 
support from the Japan Society for the
Promotion of Science (JSPS) through JSPS Research Fellowships
for Young Scientists.

\clearpage


\clearpage

\clearpage

\clearpage

%
%






\clearpage

\begin{deluxetable}{ccrccccc}
\tablecaption{Sample definitions and correlation properties.
   \label{tab:sample}}
\tablewidth{0pt}
\tablehead{
\colhead{Sample} & 
\colhead{Magnitude Range} & 
\colhead{N$_{obj}$\tablenotemark{a}}  &
\colhead{$\langle i' \rangle$\tablenotemark{b}} &
\colhead{$\langle R-i' \rangle$\tablenotemark{c}} &
\colhead{$f$\tablenotemark{d}}  & 
\colhead{A$_\omega$\tablenotemark{e}} & 
\colhead{$r_0$\tablenotemark{f} (Open,EdS)[$h^{-1}$Mpc]}
}
\startdata
whole & $i'<26$ & 1192 & 25.5 & 0.12 & 0.165  & $0.71\pm 0.26$ & 
$2.7^{+0.5}_{-0.6} (2.4^{+0.5}_{-0.5},1.6^{+0.3}_{-0.4})$\\
bright & $i'<25.5$        & 549 & 25.2 & 0.12 & 0.189  & $0.97\pm 0.57$ & 
$3.2^{+1.0}_{-1.2}    (2.9^{+0.9}_{-1.1},1.9^{+0.6}_{-0.7})$\\
faint & $25.5<i'<26$ & 643 & 25.8 & 0.10 & 0.144 & $0.56\pm0.25$ & 
$2.4^{+0.5}_{-0.7}   (2.1^{+0.5}_{-0.6},1.4^{+0.3}_{-0.4})$\\
\enddata


\tablenotetext{a}{Number of LBG candidates.}
\tablenotetext{b}{Median $i'$ magnitude of the sample.}
\tablenotetext{c}{Median $R-i'$ color of the sample.}
\tablenotetext{d}{Fraction of contaminant sources.}
\tablenotetext{e}{Best fit amplitude of power law function
for the angular correlation function. 
The slope of power law is $\beta=\gamma - 1 \equiv 0.8$ (fixed).
The reduced $\chi^2$ of the fitting is 2.1, 2.3, and 0.6 
for the whole, bright 
and faint samples, respectively.}
\tablenotetext{g}{Correlation length (comoving) 
	in a $\Lambda$-dominated universe 
	(open universe and Einstein de-Sitter universe).}

\end{deluxetable}


\end{document}